\documentclass[english,aps,prstper,reprint,showpacs,titlepage,longbibliography]{revtex4-2}   

\usepackage[T1]{fontenc}	
\usepackage[latin9]{inputenc}	
\usepackage{geometry}		
\usepackage{graphicx}
\usepackage[above,below]{placeins}	
\usepackage{times}

\usepackage{hyperref}  
\hypersetup{colorlinks=true,urlcolor=blue,citecolor=blue,linkcolor=blue}   
\usepackage{enumitem}          
\setlist{nosep}                 

\begin{document}

\begin{titlepage}

  \title{Online administration of a reasoning inventory in development}

  \author{Alexis Olsho}
  \affiliation{Department of Physics, University of Washington, Box 351560, Seattle, WA 98195-1560, USA}
  \author{Suzanne White Brahmia}
  \affiliation{Department of Physics, University of Washington, Box 351560, Seattle, WA 98195-1560, USA}
  \author{Trevor I. Smith}
  \affiliation{Department of Physics \& Astronomy and Department of STEAM Education, Rowan University, 201 Mullica Hill Rd., Glassboro, NJ 08028} 
  \author{Philip Eaton}
  \affiliation{Department of Physics, Montana State University, Bozeman, Montana 59717, USA}
  \author{Andrew Boudreaux}
  \affiliation{Department of Physics \& Astronomy, Western Washington University, 516 High St., Bellingham, WA 98225, USA}
  \author{Charlotte Zimmerman}
  \affiliation{Department of Physics, University of Washington, Box 351560, Seattle, WA 98195-1560, USA} 


  \begin{abstract}
  We are developing a new research based assessment (RBA) focused on quantitative reasoning---rather than conceptual understanding---in physics contexts. We rapidly moved administration of the RBA online in Spring 2020 due to the COVID-19 pandemic. We present our experiences with online, unproctored administration of an RBA in development to students enrolled in a large-enrollment, calculus-based, introductory physics course. We describe our attempts to adhere to best practices on a limited time frame, and present a preliminary analysis of the results, comparing results from the online administration to earlier results from in-person, proctored administration. We include discussion of online administration of multiple-choice/multiple-response (MCMR) items, which we use on the instrument as a way to probe multiple facets of student reasoning. Our initial comparison indicates little difference between online and paper administrations of the RBA, consistent with previous work by other researchers.
    \clearpage
  \end{abstract}

  \maketitle
\end{titlepage}

\section{Introduction} 
Research-based assessments (RBAs) are now widely used to examine student understanding and learning in physics instruction. In an ongoing, collaborative project, we are developing a new RBA to assess students' quantitative literacy in introductory physics contexts. Development of a valid and reliable assessment requires iterative administration and modification involving large numbers of students. For proper validation of individual items and the assessment as a whole, these administrations should occur in controlled environments. Established best practices for RBAs include in-person (proctored) administration, either on paper or electronically, for course credit but with responses not graded for correctness \cite{madsen2017}. 

RBAs can be time consuming to administer in class (both in terms of class time and in processing the data collected), resulting in interest in online administration. Research suggests that online administration of RBAs originally designed to be administered in person largely does not affect student performance \cite{bonham2008,nissen2018,wilcox2019}. However, some work suggests that online administration may result in reduced test security (with a small percentage of students copying/printing test items) and more frequent ``loss of focus''---that is, students may open other browser windows while completing the assessment \cite{nissen2018,wilcox2019}. For commonly used or well-known RBAs such as the Force Concept Inventory (FCI) or Brief Electricity and Magnetism Assessment (BEMA), students may be able to find correct answers online, though this does not prevent comparisons with results from in-person administration \cite{wilcox2019}. Decreases in student participation rates sometimes seen with online administration can be ameliorated by frequent reminders from instructors, and by following best practices similar to those for in-person administration, such as awarding credit based on completion rather than correctness. Issues of RBA security may be addressed by properly motivating the use of the RBAs with students, using a time limit, limiting the number of items students can view at one time, not giving students access to the questions outside of the online form, and requiring that students finish the survey once it has been started \cite{bonham2008,madsenphysport}.

In this paper, we share our experiences related to the rapid deployment of an online administration of an RBA in development, necessitated by the COVID-19 pandemic of 2020. In particular, we seek to answer two research questions: 1) What differences in student performance/participation are there, if any, that may be due to the difference in administration method, measured by looking at three different metrics; and 2) How does administration method affect students' response patterns for mulitple-choice/mulitple-response (MCMR) questions?

\section{RBA administration methods}
In this section, we describe the administration of our RBA under development, both on paper (in-person) and online. We discuss the circumstances under which the the online version of the RBA was administered, and our attempts to adhere to best practices in a limited timeframe. 

\subsection{Background: In-person RBA administration}

During the initial development of our RBA, we administered it to all students enrolled in the 3-quarter, large-enrollment, calculus-based introductory physics sequence at a large public university in the Pacific Northwest. We ran versions of the RBA over eight academic quarters. It was administered at the beginning of the terms, before significant instruction, thus serving as a ``pretest" for each course of the introductory sequence.

Development of a valid and reliable instrument requires regular access to a large number of students for a significant portion of instructional time. For most quarters, we were able to administer the RBA to students during recitation sessions. These sessions are typically used for required small-group activities. As students are accustomed to attending the sessions, we were able to achieve a high participation rate. This also allowed us to proctor the assessment, consistent with best-practices \cite{madsen2017}. 

Proctoring the instrument administration was resource-intensive. The assessment was administered in over 50 recitations, each 50 minutes long, during the first week of instruction. Because of the timing (during the first week of the quarter) preparing physics department TAs to proctor the assessment presented a significant challenge. 

For most in-person administrations of the assessment, students read items from a stapled packet and recorded their responses on a paper answer form as well as electronically. Our instrument includes several ``multiple-choice/multiple response'' (MCMR) items that ask students to select all answer choices they feel are appropriate. These items could not be handled by the University's multiple-choice scoring machines. 
Therefore, quarterly preparation for administration of the instrument involved not only printing the items and answer forms but also creating online surveys into which students could input their responses. Because of ongoing changes to the assessment during the development period, the stapled packets and the online surveys could not be reused. Students were asked to enter their responses online using their laptop, smartphone, or tablet if possible. Students that did not have or bring such a device with them to the class session---and so were unable to enter their answers online---were asked to indicate this on their paper answer form. After the instrument administration was finished for the quarter, a member of the research team entered those responses manually. Between 25 and 50 sets of responses were added manually each quarter.

Although we believe the methods described above resulted in high-quality data from a large number of students, they required a significant investment of time and resources. Moreover, some students misunderstood the instructions, leading research team members to spend additional time making sure the data set was complete and that students were receiving credit for their work. We began to consider online administration methods as an alternative, even exploring purchasing $\sim100$ electronic tablets through a University-based grant. In this scheme, there would be no paper version of the instrument---students would access the survey on the tablets during class, proctored by TAs or members of the research team. While this method of administration would still require significant time and effort by research team members, we believed it would be more straightforward for students than the previous procedure of entering responses online after completing the assessment on paper.

Though our focus was on in-person, proctored administration of the assessment, we began to consider whether online, unproctored administration would better support validation and wide-spread dissemination. While existing research suggests little or no significant difference in student performance between proctored and unproctored administrations of some RBAs \cite{bonham2008,nissen2018,wilcox2019}, researchers recommend that online, unproctored administration be validated separately \cite{bonham2008}. We wanted to determine whether our instrument could be administered online and unproctored by instructors who were reluctant or unable to allocate class time for administration. Moreover, though we generally have access to students during the first week of classes during scheduled recitation sessions, scheduling was difficult during academic quarters in which instruction started midweek, leading to confusion and decreased participation rates.

\subsection{Online administration}
\label{ssec:onlineAdmin}

The COVID-19 pandemic of early 2020 forced the issue. With the University moving to all online instruction, in-person administration of the assessment became impossible. Although online ``proctoring'' services exist \footnote{Many of these systems can be described more accurately as surveillance---they cannot interact with the students or provide a physical presence.}, the proctoring requirements do not align well with University policies regarding computer camera use during virtual instruction, and do not take into account possible limitations on students during such an uncertain and difficult period. 

We ran the RBA unproctored and entirely online using the University's existing survey/quiz platform. To mitigate student stress during the rapid shift to online learning, the University suggested that no graded work be required during the first week of instruction. Because we do not grade students' responses to the RBA for correctness, we decided to run the RBA, as usual, during the first week of the term in each of the three courses of the calculus-based introductory physics sequence. Because we were aware of the tendency of some students to place undue importance on such assessments, however, we presented the RBA as a low-stakes survey. 

We adhered to best practices \cite{wilcox2018,nissen2018} as much as possible: the RBA had a 50-minute time limit \footnote{The particular platform that we used to administer the online version of the instrument allows students to leave the RBA open for as long as the students want; however, after the time has expired, students cannot enter any new responses or view any more questions, though they can submit any responses already entered.}, equal to the usual class length in which the instrument was administered (we note that this is longer than it should take for students to complete the instrument); multiple reminder emails were sent to students to increase participation rate; and course credit was offered for participation, but students' responses were not graded for ``correctness.'' In addition, we constructed the online version of the instrument to discourage copying or saving of test items: each item was shown in a browser window on its own; students were not able to backtrack in the RBA \footnote{We recognize that not being able to look at questions already completed is a significant change from in-class practices, but deemed it necessary for test security purposes.} and were not shown a summary of their work or given the correct answers after completion. A video (less than 3 minutes long) embedded at the beginning of the RBA explained the purpose of the RBA and reiterated that the RBA was associated with course credit to be awarded on the basis of participation rather than the number of questions answered correctly. This is in line with best practices to discourage students from searching for answers to the items on the internet, while still motivating students to give their best efforts on the assessment \cite{bonham2008}.

Many online testing platforms will (automatically or by request) randomly order each test item's responses. We note that this does not adhere to best practices---validation of individual items only holds for the versions used during the validation process \cite{madsenphysport}. Randomizing answer choices was therefore not used to decrease cheating. Especially at the beginning of the academic quarter and with a majority of students geographically separated due to the pandemic, we believed that students were unlikely to attempt to collaborate with each other when completing the assessment. 

Because we recognized that a majority of students completing the survey for the first time would have little-to-no experience with MCMR items, we made some changes to the instrument to increase the likelihood that students would recognize that they could select multiple responses for those items. All of the MCMR items were moved to the end of the survey. After answering the last multiple-choice/single-response (MCSR) item, students saw a page with no instrument item, but rather a statement that the remaining questions on the survey might have more than one correct response, and that students should choose all answers that they feel are correct. At the top of the page for each of the remaining items (all MCMR), students saw a reminder that the question might have more than one correct response. We also prompted students to ``choose all that apply" in the question stem.

\section{Comparison of online and in-person administration: Results and discussion}
To investigate differences in student performance, we decided to compare responses from an earlier, in-person administration of the assessment to those from our online administration. We chose to use data from the in-person version of the RBA that was most similar to the online version. Of the 20 items on the assessment, only three were substantially changed between the two versions, allowing us to compare performance on the remaining 17 items.

We compare administration methods using the following metrics: 
\begin{enumerate}
    \item Participation rate.
    \item Student average score on the 17 items in common between the two versions of the instrument.
    \item The 17 items' classical test theory (CTT) \emph{difficulty} statistics.
\end{enumerate}

In addition, we compared the percentage of students choosing more than one response on the instrument's multiple-choice/multiple-response items between the two versions, to gauge whether the online instructions for the MCMR items were sufficiently clear.

The following sections present data collected in each of the three courses of the calculus-based introductory physics sequence. We refer to students in these three courses as ``C(I),'' ``C(II),'' and ``C(III),'' indicating the quarter of the instructional sequence in which the students were enrolled when completing the assessment.

\subsection{Participation rates}

Overall participation rates were similar for in-person and online administration. For in-person administration, the overall participation rate was 91\%  (93\%--92\%--89\% rate for C(I)--C(II)--C(III) students); for online administration, the overall participation rate was 90\%  (93\%--89\%--89\% for C(I)--C(II)--C(III) students). For the online administration, we counted any attempt at completing the survey as participation. (This included a small number ($<1\%$) of students who opened the survey but did not answer any of the items.)

We attribute the high participation rates on the online version to the multiple reminder emails and course web page announcements about the assessment, as well as the assignment of course credit for participating in the assessment. In addition, as in previous quarters, the assessment was associated with the weekly small-group-work recitation sessions; students were told that the survey constituted the week's work associated with the recitation session. Finally, administration of the survey during the first week, before other graded work was due, may have boosted participation, as students were not yet overly burdened with assignments.

Additionally, administering the assessment online allowed us to track the amount of time individual students took to complete it, which we were unable to do during previous in-person administrations. Although we cannot formally compare the time taken on the online version to that on the in-person versions, we do use the time data from the online administration to address student ``buy-in''---that is, whether or not students seem to take the assessment seriously. Over all three courses, the average time spent on the survey was 27.3 minutes (31.8--27.0--23.1 minutes for C(I)--C(II)--C(III) students, respectively) \footnote{This excludes the very large times (sometimes > 1000 minutes) associated with opening the survey and then ignoring it for several hours before hitting the submit button.}. Classroom observations from proctors during in-person administration suggest that students take about 40 minutes to complete the RBA in that setting. We believe the small (presumed) difference may be due to the simpler test-taking process in the online context. When completing the assessment online, students did not record their responses on paper and then enter them electronically after navigating to a website; rather, they read and responded to the items entirely online.  Time spent navigating to the website on their computer, smartphone, or tablet is not included in their time. The time-on-task data are consistent with the amount of time that we believe is necessary to read and respond to items with an appropriate amount of effort.

We did notice a small number of students in each of the courses taking ten minutes or less to complete the RBA: 5\% overall (1\%-- 3\%--11\% for C(I)--C(II)--C(III) students). Ten minutes is likely not enough time to read and consider the answer choices carefully, suggesting that these students may not have been taking the assessment as seriously as we would like. Fortunately, only for the C(III) students was the percentage of students spending less than 10 minutes a sizable fraction of the student population. Because we ran the assessment in each quarter of the 2019-2020 academic year, many of the C(III) students were seeing the assessment for the third time; we would expect these students to spend less time on the RBA due to familiarity with the material and assessment items.

\subsection{Overall student performance and item difficulty}
In this section, we compare student performance on the two administrations of the assessment, denoted ``Online'' and ``In-person''. We limit our analyses to the data collected from students enrolled in the first quarter of the calculus-based introductory physics sequence (C(I) students). We believe this is the best comparison, as these groups contain students seeing the instrument for the first time. We compared student performance on the two versions of the instrument in two ways: using the average score for a subset of 17 items in common between the two versions; and using changes in item difficulty for those 17 items. 

Average overall score and standard deviation on the subset of 17 items for Online was $8.4 \pm 3.2$ ($N=397$); In-person, it was $9.3 \pm 3.2$ ($N=326$), a percent difference of about $10\%$. While this difference is slightly larger than expected from past quarters' data, the effect is fairly small, with Cohen's $d\approx0.3$.

In addition to looking at students' scores to compare performance for the two administrations, we calculated the Classical Test Theory statistic \textit{item difficulty.} The item difficulty is the fraction of students answering each item correctly; therefore, a higher difficulty value indicates an easier question. 

\begin{figure}
\begin{center}
\includegraphics[width=.42\textwidth]{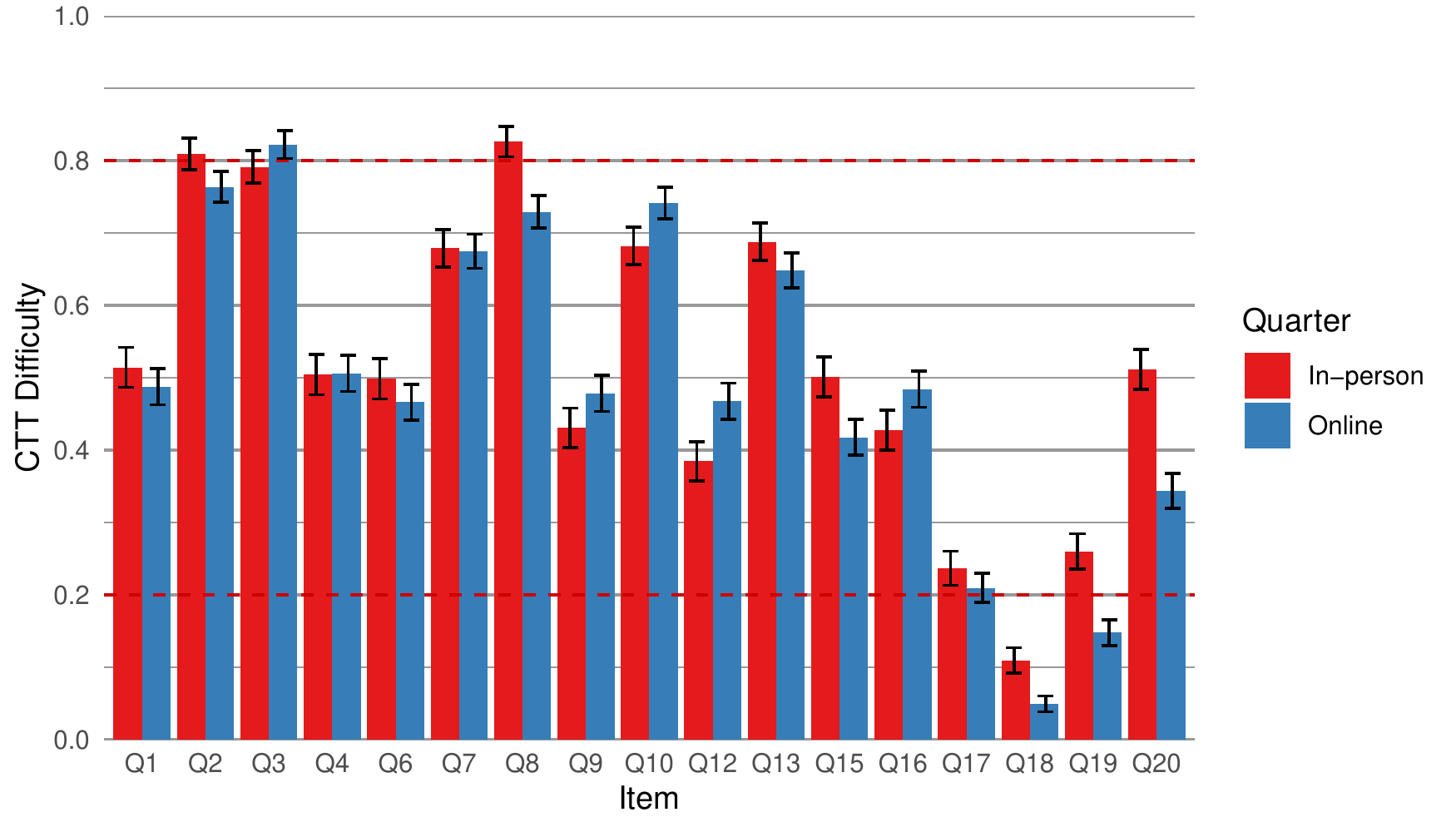}
\caption{A comparison of CTT item difficulty for 17 items from the assessment for C(I) students. Red bars represent item difficulty on the In-person administration of the assessment; blue bars are used for the Online administration. Error bars represent the standard error. Dashed lines show the upper and lower bounds for desired item difficulty.}
\label{fig:diff}
\end{center}
\end{figure}

Comparing item difficulty for the 17 common items, we found that while the average difficulty over all items in the set was not significantly different, the individual difficulty was significantly different for five items (binomial test $p < .001$). A comparison of the item difficulties is shown in Fig. \ref{fig:diff}. All five of the items had lower difficulty values for the online version of the instrument, indicating the items were more difficult for students when presented online, consistent with the lower overall score described above. Four of the five of the items (Q15, Q18, Q19, and Q20 in Fig. \ref{fig:diff}) are MCMR items; we discuss a possible explanation for the difference in section \ref{ssec:MCMR} below. We typically see large variations in the item difficulty for two of these items (Q15 and Q19), but the difficulties for those items during online administration are lower than expected from previous administrations.

\subsection{Multiple-Choice/Multiple-Response items}
\label{ssec:MCMR}

Six of the 20 items on the instrument were multiple-choice/multiple-response (MCMR), asking students to choose as many answers as they believed were correct for each item. When the instrument was administered in person, there were multiple opportunities to remind students that they could choose more than one response on these items, both in writing on the instrument itself, and also verbally by the proctor. Validation interviews suggested that multiple reminders were necessary, as this variety of question is relatively rare on the assessments typically encountered by students. We were concerned that many students would not recognize this type of question when encountering it online, especially students who had not completed the instrument previously. As noted in Section \ref{ssec:onlineAdmin} above, we made several changes to the format of the assessment to emphasize to students that they should choose more than one response for the MCMR items if appropriate. 

To assess the effectiveness of these measures, we compared the percentage of students choosing more than one response on each MCMR item, finding an increase for all MCMR items when administered online. We conclude that our measures were effective. However, as only two of the MCMR items on the RBA have more than one correct response, an increase in the number of answers chosen is not necessarily associated in an improvement in performance. Increases in the number of responses selected is generally associated with a decrease in the correct response rate, as MCMR items were scored dichotomously (i.e., an MCMR item was only counted as correct if a student selected correct answer choice(s) and did not select any of the incorrect choices). For items Q15, Q18, Q19, and Q20---the four MCMR items for which we saw significant decreases in CTT item difficulty---the fraction of students who selected more than one answer choice increased by 9\%, 22\%, 9\% and 16\%, respectively, from the In-person to the Online administration. Item 18 had two correct responses; as with the other items, there was a decrease in the item difficulty statistic and an increase in the percentage of students choosing more than one response.




\section{Discussion and future work}

In this paper we describe preliminary work toward validating an RBA in development for use with college-level introductory physics in an online, unproctored environment. Initial results tentatively suggest that students take the assessment seriously, perform at roughly the same level as for in-person administration, and are able to understand that MCMR items allow for multiple responses. To continue toward a valid and reliable \emph{online} assessment, we must learn more about how students interact with test items when using a computer or other internet-capable device, especially items for which there seems to be a significant difference in performance when administered online compared to on paper. We plan to develop an online interview protocol that may help us understand how student reasoning may change when the assessment is given in an online format. 

Although there were differences in item difficulty between the two versions of the assessment discussed, we note that most items still fall within the desired range for difficulty for first-term students, as seen in Fig. \ref{fig:diff}. The data indicate that the bulk of the difference is due to students being more willing to choose multiple responses for MCMR items. As above, we need more information about how students interact and interpret these types of questions in an online environment. Further analyses of particular answer choices on the MCMR items, going beyond dichotomous scoring, may also provide insight: for example, we are interested in changes in the percentage of students choosing both correct and incorrect responses for different administration methods. We would also like to investigate the effect of having the MCMR items interspersed with the MCSR items on the RBA as was done for the prior in-person administrations, rather than grouped together at the end.

\acknowledgments{
This work is supported by the National Science Foundation under grants No. Grants No. DUE-1832836, DUE- 1832880, DUE-1833050, DGE-1762114.
}

\bibliography{PIQLAdmin.bib}

\end{document}